\documentclass[a4paper,fleqn]{cas-dc}
\usepackage[numbers]{natbib}
\usepackage{upgreek}

\def\tsc#1{\csdef{#1}{\textsc{\lowercase{#1}}\xspace}}
\tsc{WGM}
\tsc{QE}
\tsc{EP}
\tsc{PMS}
\tsc{BEC}
\tsc{DE}

\begin{document}
\let\WriteBookmarks\relax
\def\floatpagepagefraction{1}
\def\textpagefraction{.001}
\shorttitle{}
\shortauthors{Feng et~al.~}
\let\printorcid\relax

\title [mode = title]{Influence Mechanism of Truncation on Low-Frequency Phase Measurement}                      

\author[1]{Yujie Feng}
\credit{Conceptualization, Investigation, Writing - original draft, Software}

\author[1]{Yuanze Jiang}
\credit{Investigation, Conceptualization}

\author[1]{Liuyang Chen}
\credit{Formal analysis, Data curation}

\author[1]{Haifeng Chen}
\credit{Software}

\author[1]{Yurong Liang}
\credit{Funding acquisition, Supervision, Writing – review \& editing}
\cormark[1]

\affiliation[1]{organization={MOE Key Laboratory of Fundamental Physical Quantities Measurement and Hubei Key Laboratory of Gravitation and Quantum Physics, PGMF, School of Physics, Huazhong University of Science and Technology},
                city={Wuhan},
                postcode={430074}, 
                country={People’s Republic of China}}

\cortext[cor1]{Corresponding author: liangyurong20@hust.edu.cn}

\begin{abstract}
Driven by advances in electronic technology, modern digital phasemeters have significantly improved in integration and functionality, enabling real-time measurement and analysis of dynamic signals. High-precision phase measurement is closely associated with the quantization process. This paper specifically analyzes the white and non-white noise characteristics associated with the quantization errors of phase truncation in digital phasemeters. The error can be considered white noise under specific conditions, which power correlates with the resolution of quantizer and is uniformly distributed within the Nyquist frequency. However, when the signal frequency and sampling frequency are close to an integer multiple, the non-white noise caused by truncation can result in low-frequency phase noise. Additionally, artifacts may induce nonlinear phase errors. Introducing Gaussian dither synthesized by LFSRs can smooth the truncation process, thereby mitigating its impacts on phase measurement. The results indicate that for a 10 MHz signal under test, the noise floor of the phasemeter exceeds the requirement from 2 mHz to 0.1 Hz due to the integer multiple. After adding dither, the phase noise was optimized by 9.5 dB at 10 mHz, achieving the requirement of 1.3 $\rm{\upmu rad/Hz^{1/2}} \cdot \rm{NSF}$ from 0.1 mHz to 1 Hz in space gravitational wave detection. This demonstrates that adding dither can effectively suppress the low-frequency phase noise caused by truncation.
\end{abstract}

\begin{keywords}
Quantization \sep Truncation \sep Phase measurement \sep Low-frequency noise
\end{keywords}

\maketitle

\section{Introduction}
Modern digital phasemeters have significantly enhanced both integration and functionality, driven by advancements in field-programmable gate array (FPGA) and digital signal processing (DSP) technologies. Digital phasemeters incorporate multiple functions, including frequency counters and amplitude meters, to address diverse testing requirements. They offer real-time measurement and analysis capabilities, which makes them particularly suitable for evaluating dynamic signals. In communication systems, phasemeters are utilized for phase noise and jitter analysis; in radar and navigation systems, they enable high-precision measurements of distance and speed. In addition, phasemeters find extensive applications in electronic testing, motor control, high-energy physics, and astronomy. The frequency synthesizer, a critical component of digital phasemeters, employs direct digital synthesis (DDS) technology to generate digital waveforms at specified frequencies. It provides precisely matched quadrature signals to satisfy the stringent demands for resolution and bandwidth in phasemeters~\cite{Foster2006Integrated}.

High-precision phase measurement is a key technique in space gravitational wave detection~\cite{Ming2020Ultraprecision}. The technical approach of current mainstream projects is similar. A formation of three satellites is employed to construct a heterodyne laser interferometer with arm lengths of one million kilometers in space~\cite{Karsten2003LISA,Luo2016TianQin,Hu2017Taiji,Gong2021Concepts,Jennrich2001Demonstration}. As gravitational waves traverse the system, changes in space-time between the test masses are converted into phase shifts in the inter-satellite interferometers. Due to the Doppler shifts caused by the periodic relative motion between satellites, the digital phasemeter must extract scientific signals in the frequency range of 0.1 mHz to 1 Hz from a carrier frequency of 5 to 25 MHz, and the phase measurement resolution needs to reach the order of $\rm{\upmu rad/Hz^{1/2}}$.

Early phase measurement included zero-crossing and IQ demodulation methods~\cite{Luo2020brief,Shaddock2006Overview}, but these approaches could not perform real-time phase measurement for swept-frequency signals. Subsequently, the Jet Propulsion Laboratory and the Albert Einstein Institute began developing FPGA-based digital phasemeters~\cite{Wand2006LISA,Gerberding2013Phasemeter}. Utilizing a digital phase-locked loop (DPLL), these phasemeters can accurately demodulate scientific signals from carriers with continuously changing frequencies, becoming a mainstream scheme for phase measurement in gravitational wave detection. After more than a decade of development, digital phasemeters have achieved significant performance improvements~\cite{Gerberding2014Phase,Gerberding2015Readout,Liang2015Note,Bode2024Noise,Zhang2024Multi,Feng2024Utilizing}. 
However, it has been observed that when the signal frequency and sampling frequency are close to an integer multiple, the risk of low-frequency phase noise increases. To avoid this noise during testing, Non-integer ratio frequencies, such as 10.3 MHz, are often employed. In practical applications, the input signal is a high-speed carrier with a frequency sweep rate of $\rm{Hz/s}$, inevitably passing through integer frequency ratios. The low-frequency noise of the phase will increase.

Prior studies have focused on the metrological properties of DDS. Tian et al.~\cite{Tian2009DDFS,Tian2010Spurious} investigated the influence of amplitude quantization and phase truncation on output spurs signals in DDS, describing the spurs characteristics of generated signals through theoretical derivation and simulation results. Rybin et al.~\cite{Rybin2017Basic} noted that waveforms generated by DDS are unsatisfactory in terms of metrology, as they carry amplitude errors, period errors, and generate spurs signals, resulting in a complex signal spectrum. However, these studies did not elucidate the mechanisms behind the increase in low-frequency phase noise. High-precision phase measurement is closely related to the quantization process. These errors can be treated as white noise under certain conditions. However, the output waveform of the synthesizer is affected by discrete phase accumulation and phase truncation processes. When the clock and signal frequencies are in an integer multiple, this can lead to higher-level spurs appearing at specific frequencies. Consequently, the error exhibits non-white noise characteristics within the Nyquist bandwidth, resulting in low-frequency phase noise.

This paper investigates the influence mechanisms and suppression methods of truncation on low-frequency phase measurement. Firstly, it introduces the definition of phase noise and the basic principles of digital phasemeters. Secondly, we analyze the white and non-white noise characteristics associated with the quantization errors of phase truncation in digital phasemeters. Finally, a scheme for reducing phase noise by adding Gaussian dither is introduced, and the experimental results show that when the signal to be measured is 10 MHz, the noise floor of the phasemeter exceeds the requirement from 2 mHz to 0.1 Hz due to the integer multiple, and an optimization of 9.5 dB can be achieved by adding dither. The requirement for phase measurement in space gravitational wave detection is satisfied. It is demonstrated that the effect of truncation on phase can be effectively mitigated by adding dither.
\begin{figure*}
	\centering
	\includegraphics[width=15.5cm]{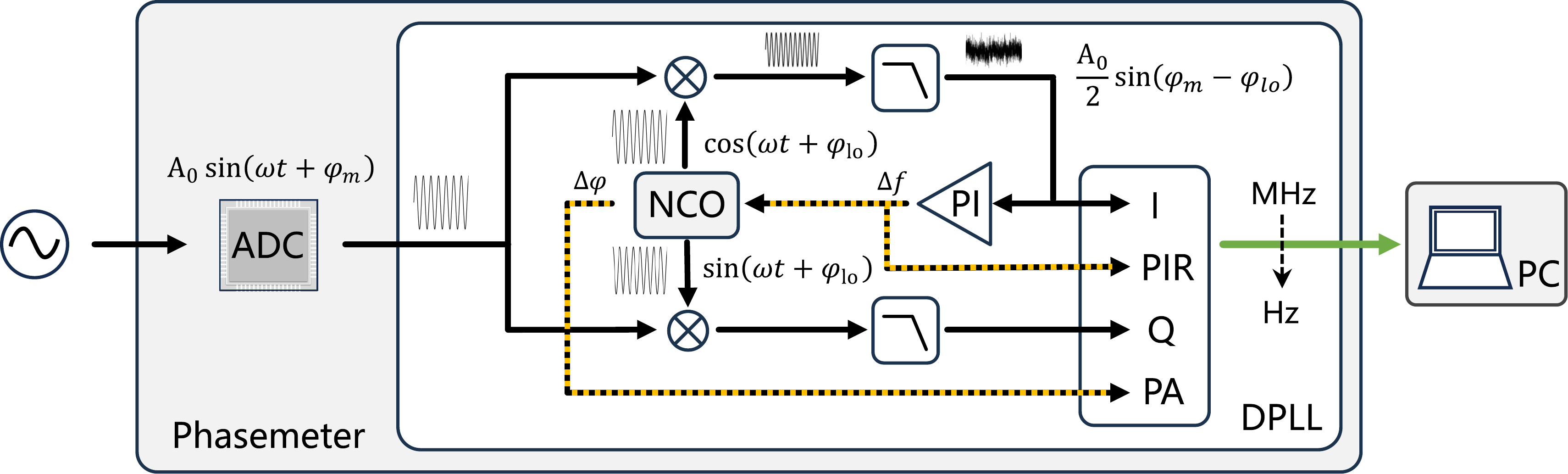}
	\caption{Schematic diagram of a DPLL-Based Digital Phasemeter~\cite{Yamamoto2023Intersatellite}. The signal digitized by the ADC enters the FPGA for phase demodulation. The NCO serves as the frequency synthesizer. The PA adds its output to the PIR to generate reference signals. Phase detection is performed through mixing and filtering. When the loop is locked, the PA and PIR can directly indicate the phase changes of the input signal. The phase data received by the PC is downsampled to conserve hardware resources.}
	\label{FIG:1}
\end{figure*}

\section{Phase Measurement Principle}\label{section2}
Phase noise, denoted by ${S_{\varphi}}(f)$, is defined as the single sideband power spectral density of a signal's random phase jitter. The industry typically uses $L(f)$ to denote phase noise, as defined by IEEE Standard 1139~\cite{IEEE2009Std}:
\begin{equation}
    L(f) = \frac{1}{2}S_{\varphi}(f),\quad[L(f)]=\rm{dBc/Hz}
\end{equation}
Phase noise represents the ratio of single sideband noise power within a 1 Hz bandwidth at an offset frequency ${\Delta}f$ to the carrier power. This definition is analogous to the carrier-to-noise ratio ($\rm{C/N_0}$), and phase noise can be determined by calculating the ratio of carrier power to noise power. An increase in the spectral linewidth of the carrier signal results in a corresponding increase in phase noise.

The digital phasemeter is well suited for dynamic signal evaluation, possessing the capability to measure and analyze signals in real time. \textcolor{blue}{Fig.}\ref{FIG:1} illustrates the principle of a DPLL-based digital phasemeter. When the clock and signal frequencies are non-integer ratio frequencies, the in-phase (I) and quadrature-phase (Q) signals can be expressed as:
\begin{equation}
\begin{split}
    {\rm{I}} = {\rm{A_0}} \sin \left( \omega t + \varphi_m \right) \cos \left( \omega t + \varphi_{lo} \right) \approx \frac{1}{2} \rm{A_0} \sin \Delta \varphi\\
    {\rm{Q}} = {\rm{A_0}} \sin \left( \omega t + \varphi_m \right) \sin \left( \omega t + \varphi_{lo} \right) \approx \frac{1}{2} \rm{A_0} \cos \Delta \varphi
\end{split}
\end{equation}
In the above equations, ${\Delta}{\varphi}={\varphi}_m-{\varphi}_{lo}$. When the loop is locked, the phase accumulator (PA) and phase increment register (PIR) can directly show the phase changes of the input signal. The theoretical model of a linearized DPLL is well-established and the transfer functions can be expressed as follows:
\begin{equation}
    H(z) = 2^{M+N-3}\frac{V_{\rm{in}}}{V_{\rm{F}}}\left(K_p + \frac{K_iT}{1-z^{-1}}\right)\frac{f_{\rm{s}}}{2^A}\frac{2{\pi}T}{1-z^{-1}}     
\end{equation}
Where $N$, $M$, and $A$ represent the bit widths of the ADC, numerically controlled oscillator (NCO), and PIR, respectively. $V_{\rm{F}}$ represents the full-scale range of the ADC, and $T$ represents the reciprocal of the sample rate $f_{\rm{s}}$, indicating the integration time.

\section{Quantization Error of Phase Truncation}
Analog circuit noise sources, such as thermal noise, shot noise, and flicker noise, originate from physical phenomena. In contrast, digital noise sources, including quantization noise and rounding noise, arise from the computational operations. Quantization operations approximate analog amplitudes with digital representations, inherently introducing errors known as quantization noise. Analyzing the characteristics of quantization noise is crucial for understanding the digital signal processing workflow~\cite{Levy2020Random}.

Quantization errors are often analyzed as white noise, which necessitates specific conditions~\cite{Oppenheim2009Discrete}. Bennett~\cite{Bennett1984Spectra} observed that for sufficiently small discretization steps, quantization noise can be approximated as uniformly distributed and white. Widrow~\cite{Widrow1956Study,Widrow1961Statistical} identified the exact conditions required for these properties to hold, and later Sripad and Snyder~\cite{Sripad1977necessary} proposed the necessary and sufficient conditions for their establishment. In summary, when quantization noise can be regarded as white, the following conditions must be met:

\begin{enumerate}
\itemsep = 0pt
\item The signal $x(t)$ approaches different levels with equal probability.
\item The quantization step Q is small and uniform enough.
\item The  error $e(n)$ is independent of the signal $x(t)$.
\end{enumerate}  
The white and non-white noise characteristics of quantization errors will be discussed in this section.

\subsection{White Noise}
\begin{figure}[ht]
	\centering
	\includegraphics[width=7.5cm]{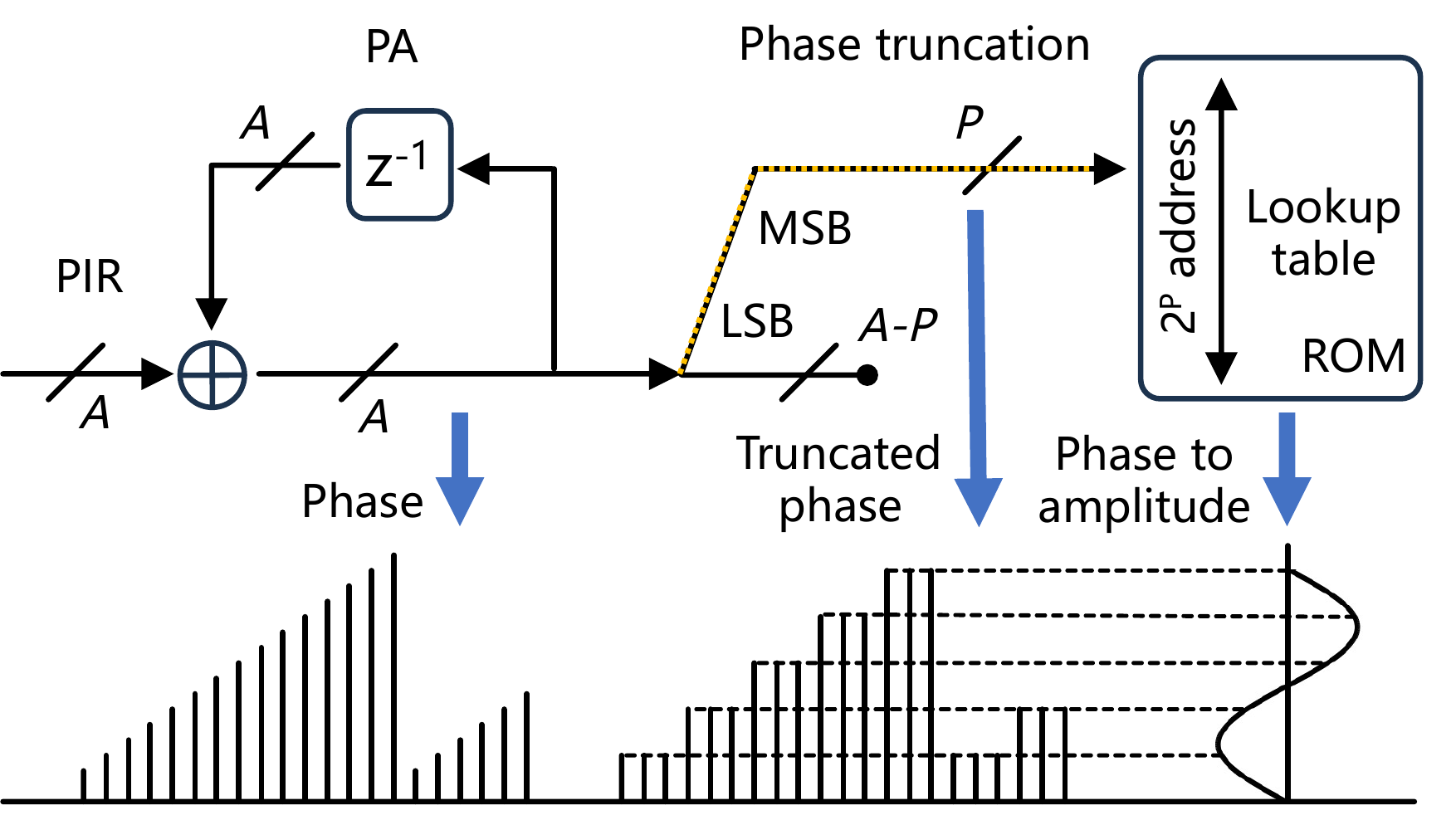}
	\caption{The schematic diagram of phase truncation in the NCO illustrates the phase-amplitude conversion process. During this process, the PA adds its output to the PIR. The most significant bits (MSB) of the PA are then truncated and used as addresses in a Look-Up Table (LUT) to generate reference signals.}
	\label{FIG:3}
\end{figure}

Phase truncation is an important process in NCO, allowing for resource savings while still meeting measurement requirements. If all 64 bits PA are utilized for phase-to-amplitude conversion, $2^{64}$ entries would need to be stored in the lookup table (LUT). and the LUT would require approximately $2\times10^7$ TB of memory. As shown in \textcolor{blue}{Fig.}\ref{FIG:3}, the MSB of the PA output is used to provide phase. By using 12 bits, the lower 52 bits are truncated, resulting in a memory requirement of only 4 kB.

In the phase-amplitude conversion process, quantization errors introduced by truncation result in periodic amplitude errors. Assuming that the period is sufficiently large, the error traverses each value before repeating. Therefore, truncation errors are uniformly distributed within the Nyquist bandwidth and exhibit white noise characteristics. Let $P$ represent the bit width after truncation. The quantization noise caused by truncation is expressed as:
\begin{equation}
    \widetilde{n}_{\rm{trun}} = \frac{2^{-P}}{\sqrt{6f_{\rm{s}}}},\quad[\widetilde{n}_{\rm{trun}}] = \rm{1/Hz^{1/2}}
\end{equation}

\begin{figure}[ht]
	\centering
	\includegraphics[width=5.5cm]{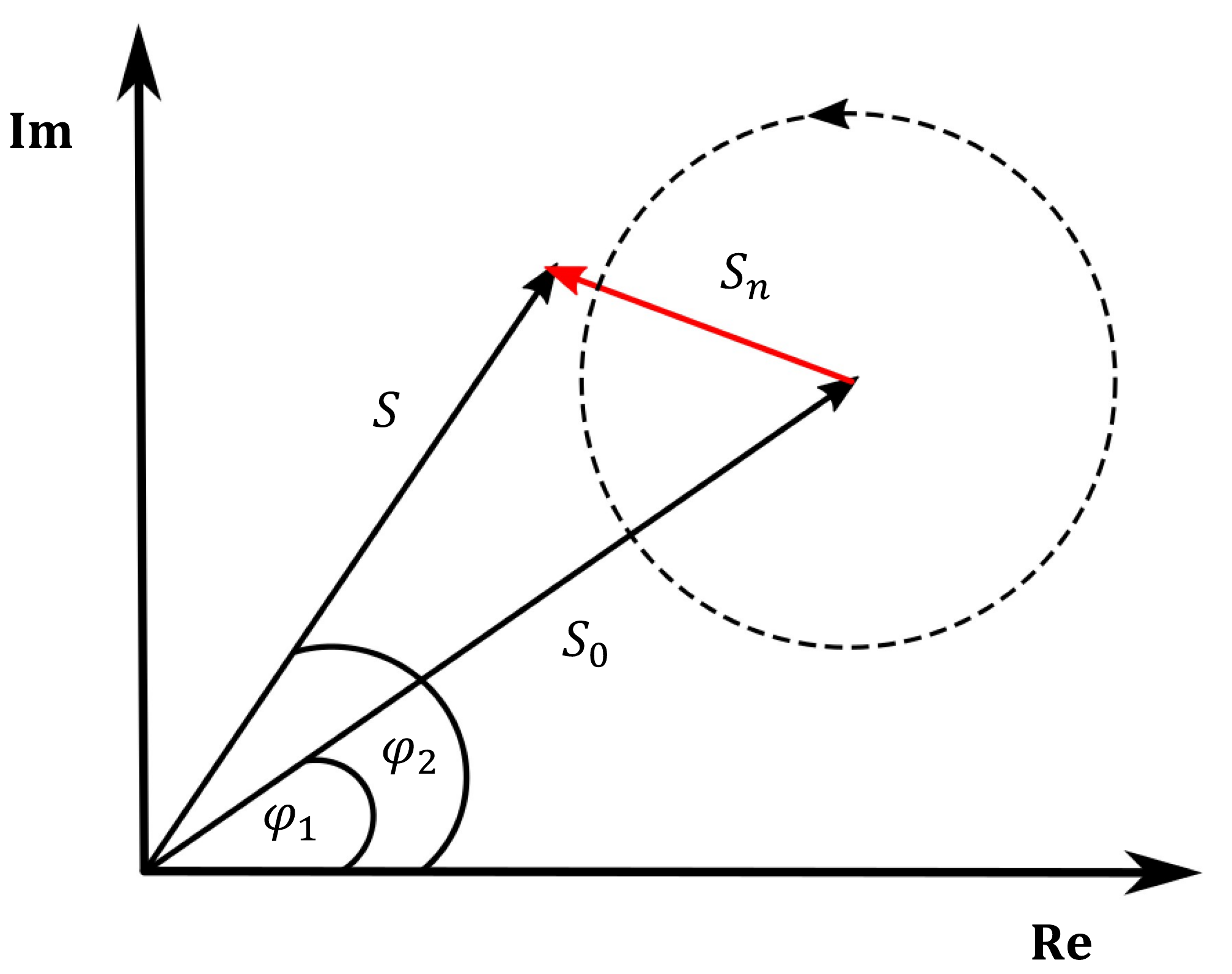}
	\caption{Complex phasors are used to describe the phase coupling mechanism of additive noise. Quantization noise phasor $S_n$ cause a phase deviation in the signal phasor $S$.}
	\label{FIG:2}
\end{figure}

This white noise affects phase measurement. Complex phasors are utilized to describe the general coupling mechanism of an additive noise~\cite{Wissel2022Relative}, as shown in \textcolor{blue}{Fig.}\ref{FIG:2}, where $S_0$ is the signal phasor, $S_n$ is the small noise phasor, resulting in the phasor $S$. The angle between $S_n$ and $S_0$ is ${\pi}/2$ when the influence is maximal, the error can be expressed as: 
\begin{equation}
    {\varphi}_{\rm{err}} = {\varphi}_2-{\varphi}_1\approx{\rm{tan}}( {\varphi}_{\rm{err}})\approx\frac{|S_n|}{|S_0|}
\end{equation}
So the phase noise caused by truncation is described by:
\begin{equation}
    \widetilde{\rm{\varphi}}_{\rm{trun}} = \frac{2^{-P}}{\sqrt{6f_{\rm{s}}}},\quad[\widetilde{\rm{\varphi}}_{\rm{trun}}] = \rm{rad/Hz^{1/2}}
\end{equation}
When $P$=12 bits, $f_{\rm{s}}$=80 MHz, the $\widetilde{\rm{\varphi}}_{\rm{trun}}$ is about 0.03 $\rm{\upmu rad/Hz^{1/2}}$, which has minimal impacts on phase measurement and can be disregarded.

\subsection{Non-white Noise}

\begin{figure*}
	\centering
	\includegraphics[width=12cm]{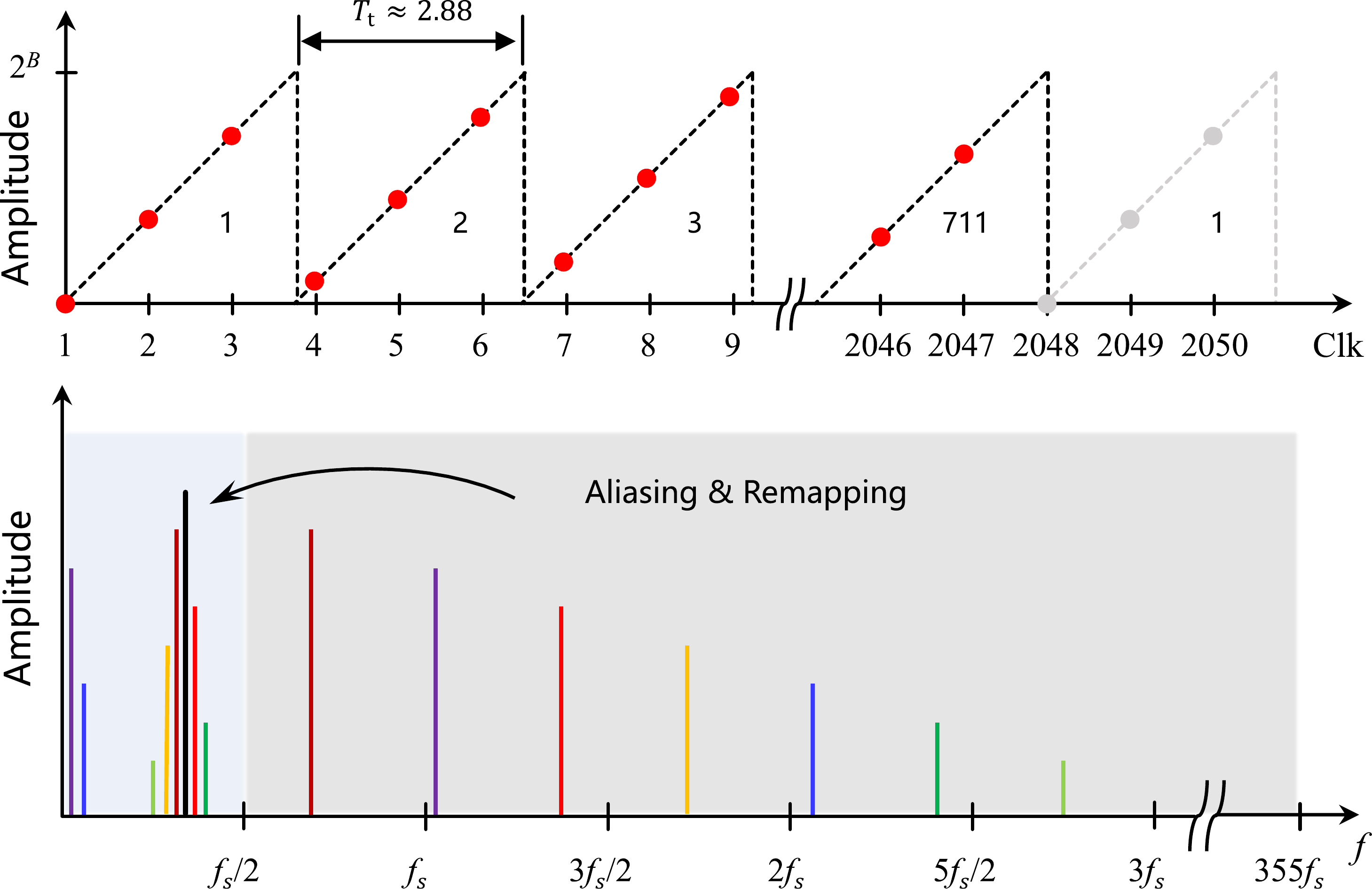}
	\caption{
    Overflow behavior of truncation word accumulator (Top) and spectrum of truncation spurs (Bottom). The overflow behavior exhibits a sawtooth waveform with a period of $T_t\approx$2.88 clock cycles. Additionally, the complete sequence of truncation word values repeats after a period of GRR=2048 clock cycles. Since the truncation word behavior is periodic in the time domain, its Fourier transform is periodic in the frequency domain. Therefore, 1,024 discrete frequencies constitute the truncation spurs. When $f>f_{\rm{s}}/2$, these spurs are aliased into the Nyquist bandwidth. The results illustrates the spectral lines that remap the actual output, representing the truncated spurs spectrum of the NCO output. However, this figure only demonstrates concepts and is not entirely accurate; it only shows the frequency range within 3$f_{\rm{s}}$. The spectrum actually spans 355$f_{\rm{s}}$, and the remapped lines are significantly more numerous.}
	\label{FIG:4}
\end{figure*}

When the signal frequency and sampling frequency are close to an integer multiple, $e(n)$ and $x(t)$ are no longer independent, and the error exhibits non-white noise characteristics. This phenomenon primarily occurs during the phase truncation process. Precisely analyzing the distribution of truncation-induced spurs is quite complex~\cite{Nicholas1987Analysis}. This process will be illustrated through a specific and illustrative example.

\begin{figure}[ht]
	\centering
	\includegraphics[width=8cm]{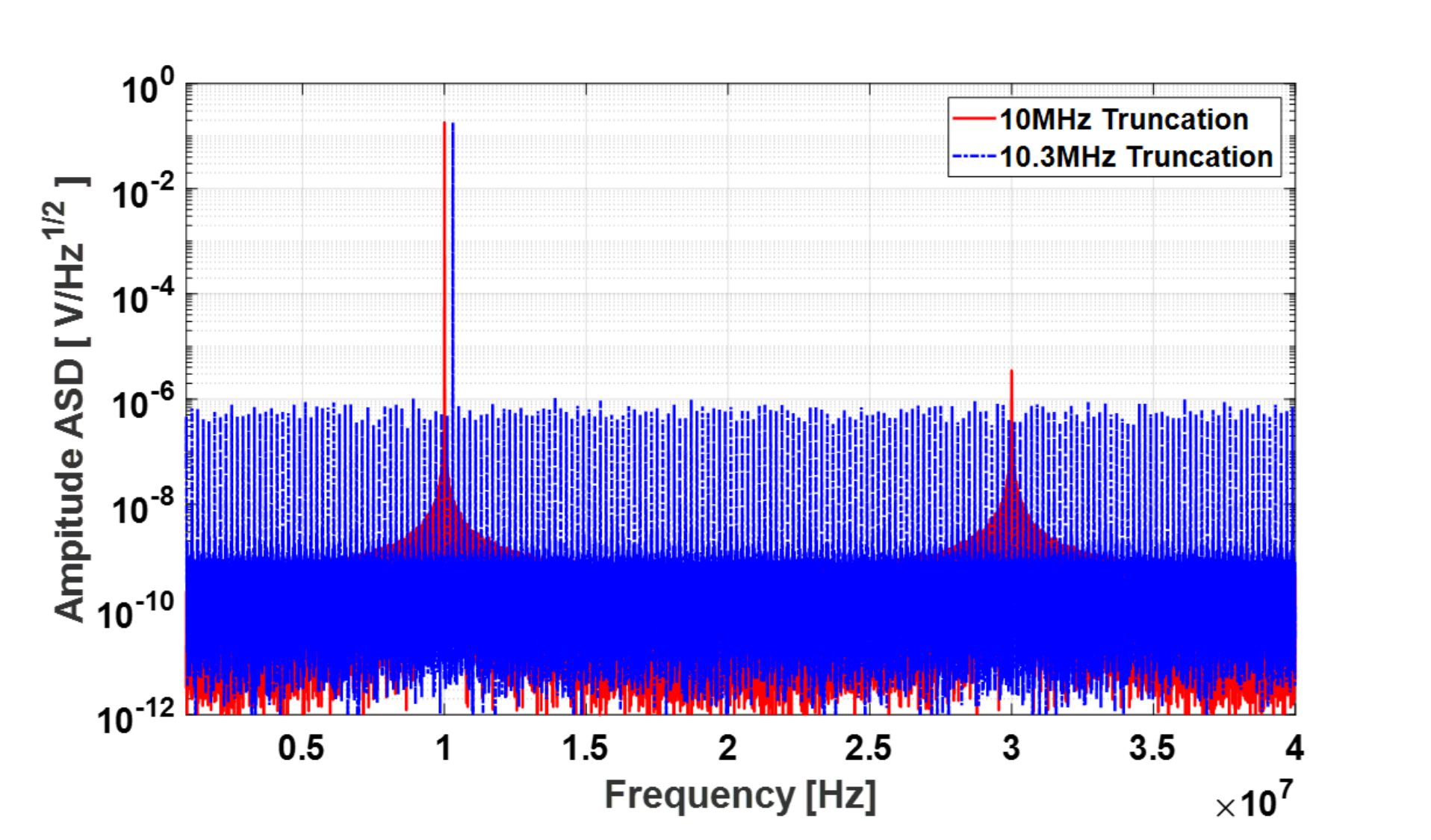}
	\caption{Non-white noise characteristics are observed in the simulation results. When the signal frequency is 10.3 MHz, the spurs generated by truncation are evenly distributed across the Nyquist bandwidth, exhibiting the statistical characteristics of white noise. Conversely, when the frequency is about 10 MHz, the spectral lines remapped by aliasing are concentrated at 10 MHz and 30 MHz. This process results in an increase in spectral linewidth at 10 MHz and the generation of artifacts at 30 MHz.}
	\label{FIG:5}
\end{figure}

In successive clock cycles, the PIR is added to the current content of the PA. After a certain period, the PA returns to its initial PIR value and begins to repeat the cycle, as depicted in \textcolor{blue}{Fig.}\ref{FIG:4}. The number of this clock cycles is called the grand repetition rate (GRR):
\begin{equation}
{\rm{GRR}} =  2^A/~{\rm{GCD}(PIR},2^A)
\end{equation}
Where GCD is the greatest common divisor. The output of NCO is a combination of ideal signal and error signal. The error signal caused by truncation is the source of spurs, so we focus on analyzing the truncated part, which represents a $B$ bits PA. The equivalent tuning word (ETW) is:
\begin{equation}
{\rm{ETW}} = {\rm{PIR~mod}}~2^B
\end{equation}
When $B$=12 bits, the ETW=2,674 is substituted into Equation (8) as the new PIR, resulting in GRR=2,048. The behavior of the truncation word determines the characteristics of the spurs. Consequently, the overflow period $T_{\rm{t}}\approx$ 2.88 clock cycles can be got by:
\begin{equation}
T_{\rm{t}}=\left\{\begin{array}{cr}
\frac{2^B}{\rm{ETW}}, & {\rm{ETW}}<2^{B-1} \\
\frac{2^B}{2^B-\rm{ETW}}, & {\rm{ETW}} \geq 2^{B-1}
\end{array}\right.
\end{equation}
So this process can be described by a sawtooth wave with a fundamental frequency of 0.35$f_{\rm{s}}$. In summary, the spurs span the frequency range of 355$f_{\rm{s}}$. This leads to spurs aliasing into the Nyquist bandwidth. The results of simulation and spectrum analysis are shown in \textcolor{blue}{Fig.}\ref{FIG:5}.

When the signal frequency is about 10 MHz, the spectral lines remapped by aliasing are concentrated at 10 MHz and 30 MHz. This process results in an increase in the spectral linewidth at 10 MHz and the generation of artifacts at 30 MHz. On one hand, from the definition of phase noise in Section \ref{section2}, it can be observed that $\rm{C/N_0}$ of the output signal deteriorates with an increase in the spectral linewidth. Consequently, the truncation spurs reflect the non-white noise characteristics in the phase measurement. On the other hand, considering the artifacts, the reference signal $v_{\rm{ref}}$ generated by the NCO includes:
\begin{equation}
v_{\rm{ref}} = \sin \left(\omega t+\varphi_{lo}\right) + {\rm{A_1}} \sin \left(3 \omega t+\varphi_{lo}\right)
\end{equation}
The sum of $3\omega$ and $\omega$ after phase detection is near the Nyquist frequency, and it is remapped near DC after aliasing, then retained after low-pass filtering. Therefore, the phase change is:
\begin{equation}
\Delta \varphi=\arctan \left(\frac{\sin \left(\varphi_m-\varphi_{l o}\right)+\frac{\rm{A_1}}{\rm{A_0}} \sin \left(\varphi_m+\varphi_{l o}\right)}{\cos \left(\varphi_m-\varphi_{l o}\right)-\frac{\rm{A_1}}{\rm{A_0}} \cos \left(\varphi_m+\varphi_{l o}\right)}\right)
\end{equation}
Analyze $G = {\rm{A_1}}/{\rm{A_0}}$  as a perturbation. Using a first-order Taylor expansion:
\begin{equation}
\begin{aligned}
\left. \frac{d \varphi}{d G} \right|_{G=0} &= 
\frac{\cos \left( \varphi_m - \varphi_{lo} \right) \sin \left( \varphi_m + \varphi_{lo} \right)}
{\cos^2 \left( \varphi_m - \varphi_{lo} \right) + \sin^2 \left( \varphi_m - \varphi_{lo} \right)} \\
&+\frac{\sin \left( \varphi_m - \varphi_{lo} \right) \cos \left( \varphi_m + \varphi_{lo} \right)}
{\cos^2 \left( \varphi_m - \varphi_{lo} \right) + \sin^2 \left( \varphi_m - \varphi_{lo} \right)} \\
&= \sin \left( 2 \varphi_m \right)
\end{aligned}
\end{equation}
After considering the artifacts, phase can be expressed as:
\begin{equation}
\Delta \varphi_t = \Delta \varphi+ G \sin \left(2 \varphi_{m}\right)
\end{equation}

\begin{figure}[ht]
	\centering
	\includegraphics[width=8cm]{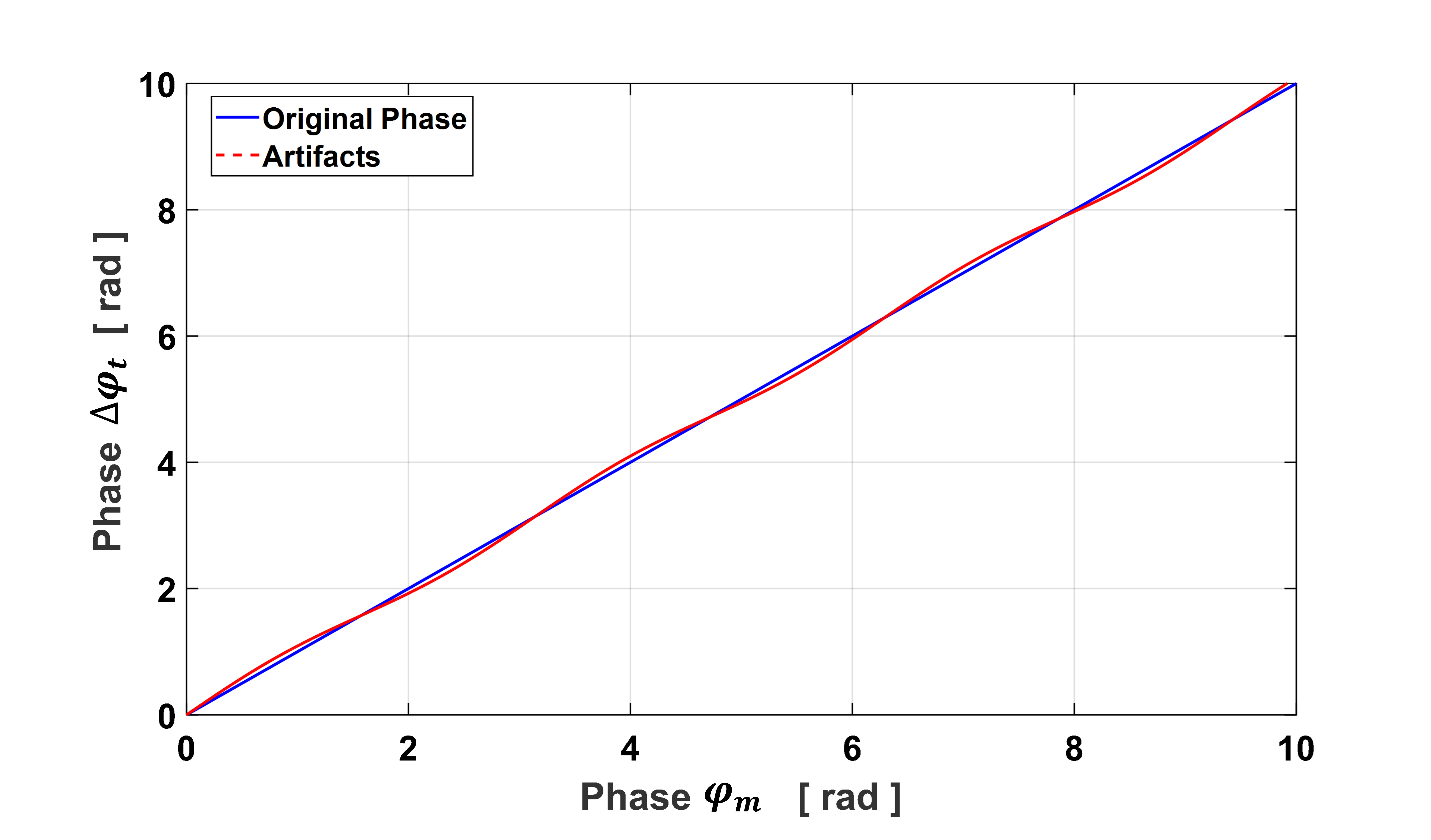}
	\caption{The presence of specific artifacts introduces a nonlinear effect on the phase. The solid blue curve illustrates the linear relationship between the phase measurement results and the true phase change when $G$=0. In contrast, the dashed red line represents the nonlinear phase measurement results, taking into account the underlying artifacts.}
	\label{FIG:6}
\end{figure}

The remapped spectral lines are evenly distributed across the Nyquist bandwidth when the clock and signal frequencies are non-integer ratio frequencies. However, as the two frequencies approach integer multiple, the remapped spectral lines tend to concentrate at specific frequency points, resulting in the so-called artifacts. Equation (13) demonstrates that the presence of a specific artifact introduces a nonlinear effect on the measured phase value, as shown in \textcolor{blue}{Fig.}\ref{FIG:6}. This effect is influenced by the initial phase of the signal being measured and the amplitude of the artifacts, which ultimately causes the appearance of white noise characteristics in the phase measurement.

\section{Dither Synthesis and Experiment}
Quantization error is a deterministic function of the input signal~\cite{Levy2020Random}, which has a significant influence on phase measurement when the clock and signal frequencies are non-integer ratio frequencies. Dithering, a common technique, introduces a small amount of noise to the input signal to prevent the error from "sticking" to a specific value, thus reducing the correlation between the quantization error and the input signal~\cite{Widrow2008Quantization}. In practice, synthetic white noise can be deliberately added before quantization to smooth the truncation process, thereby mitigating the impacts on phase measurement.

\begin{figure}[ht]
	\centering
	\includegraphics[width=8cm]{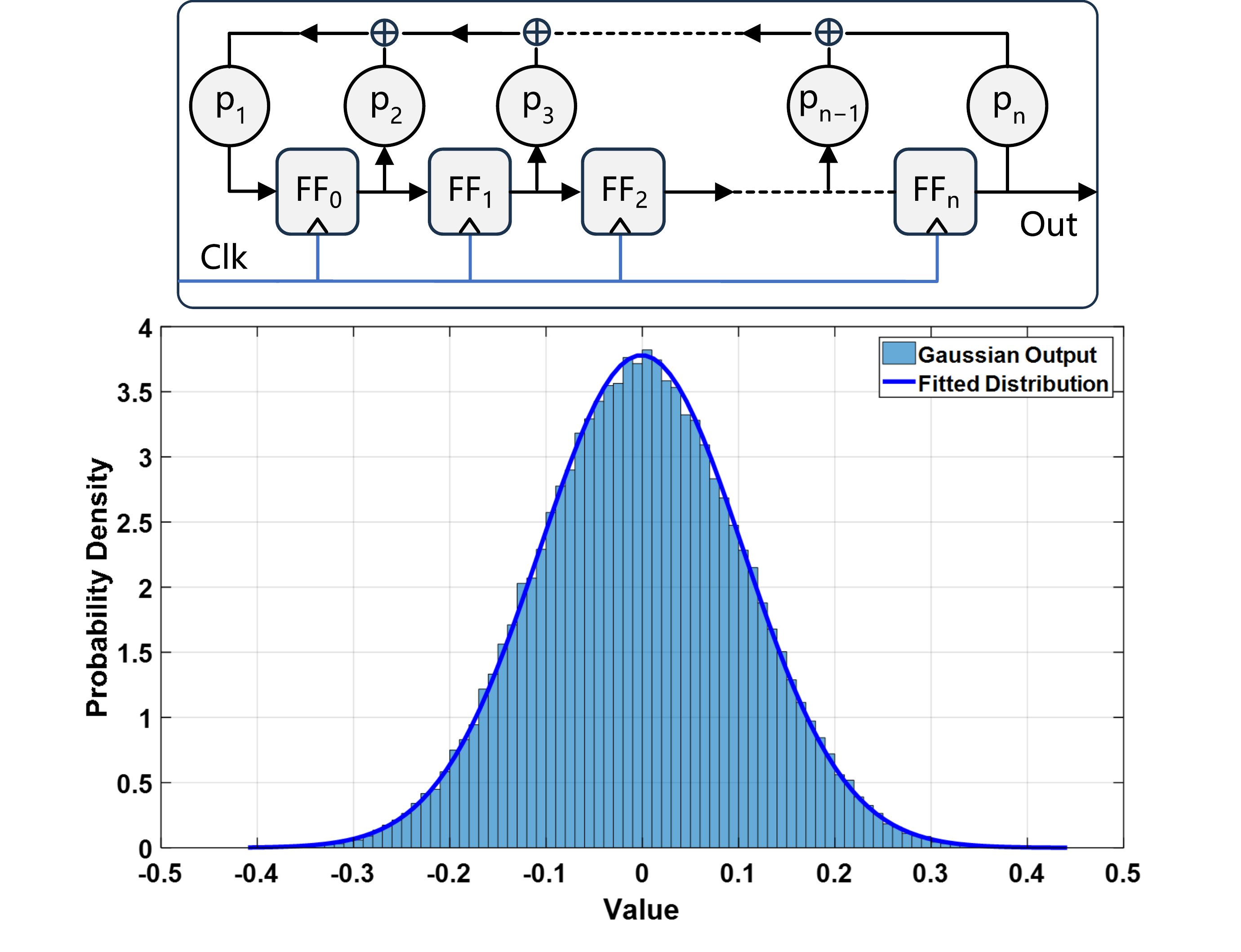}
	\caption{The LFSR consists of several registers and XOR gates and can be conveniently implemented using Verilog. The schematic diagram of the Fibonacci LFSR is shown at the top. A Gaussian dither histogram with its fitting curve is presented at the bottom, with the overall probability normalized to 1.}
	\label{FIG:7}
\end{figure}

Gaussian white noise can be effectively synthesized using the central limit theorem (CLT) and linear feedback shift registers (LFSR). This synthesis is ideal for dither analysis, as the CLT asserts that multiple random samples from a uniform distribution yield a normal distribution for independent, identically distributed variables. In digital circuits, LFSRs are instrumental in generating pseudo-random, uniformly distributed numbers. Specifically, when a primitive polynomial determines the coefficients ${\rm{P_n}}$, the LFSR achieves its maximum period $T_{\rm{max}} = 2^{n-1}/f_{\rm{s}}$, where $n$ is the bit width of LFSR. To prevent the synthesized noise from repeating within the required 0.1 mHz frequency band, the period of LFSR must exceed 10,000 seconds. If $f_{\rm{s}}$=80 MHz, $n$ must be greater than 39.5 bits. As illustrated in \textcolor{blue}{Fig.}\ref{FIG:7}, the Gaussian dither histograms generated by Verilog align with the anticipated normal distribution. 

\begin{figure*}
	\centering
	\includegraphics[width=16cm]{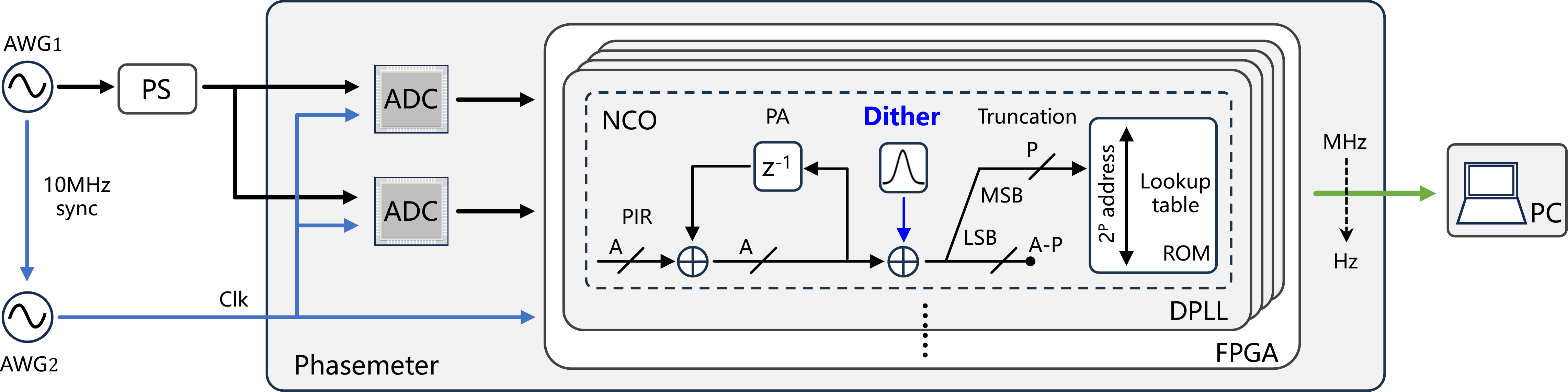}
	\caption{The signal is generated by AWG1, an arbitrary waveform generator Keysight 33622A. This signal is evenly split into two input paths for phase measurement using a Mini-Circuits ZMSCJ-2-2 power splitter. The prototype phase meter architecture is implemented on the Terasic DE2-115 commercial FPGA platform. The signal to be measured is digitized by ADI AD9254 and fed into the FPGA for zero measurements. The phase meter clock is derived from AWG2, which is synchronized with AWG1. This synchronization ensures that the reference signal generated by the NCO is precisely aligned with the sampling clock of the ADC, facilitating the accurate acquisition of integer frequency signals and supporting subsequent experimental validation.}
	\label{FIG:9}
\end{figure*}

\begin{figure}[ht]
	\centering
	\includegraphics[width=8cm]{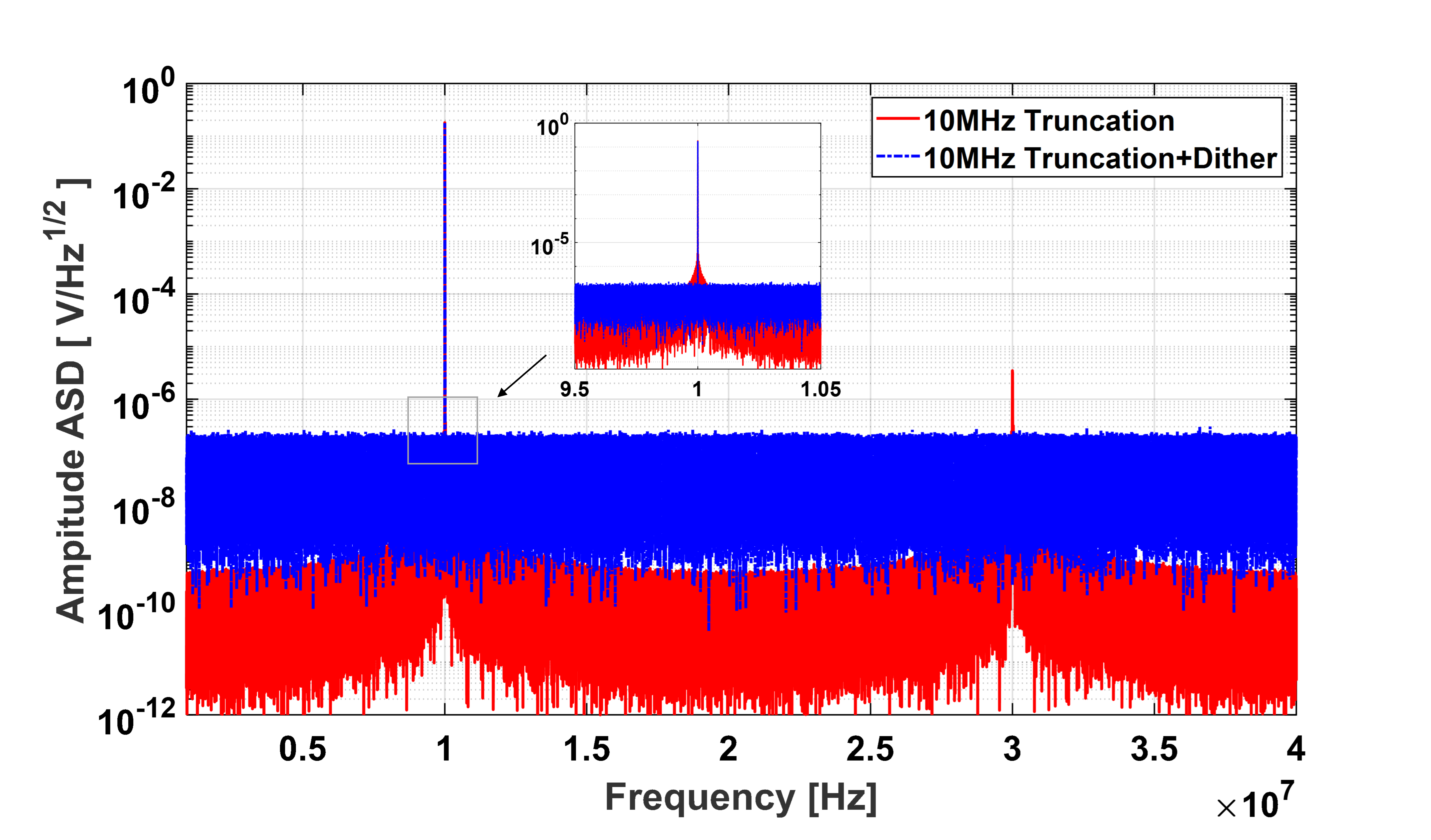}
	\caption{The spectrum analysis of the NCO output simulation results, after adding dither, reveals the following observations: the spectral linewidth of the 10 MHz reference signal has been optimized, leading to enhanced signal clarity and stability. Additionally, any artifacts at 30 MHz have been effectively mitigated, ensuring a cleaner output spectrum.}
	\label{FIG:8}
\end{figure}

Furthermore, the simulation results of adding Gaussian dither show in \textcolor{blue}{Fig.}\ref{FIG:8}. It improves the quality of the NCO output effectively. The experimental results provide further evidence to demonstrate their practical applicability. In the experiment, phase noise floor was evaluated using a differential test scheme, as illustrated in \textcolor{blue}{Fig.}\ref{FIG:9}. Two signals with frequencies of 10 MHz and 10.3 MHz were measured for more than 10,000 seconds. Compared to the 10.3 MHz signal, the 10 MHz phase spectrum clearly reveals the influence of non-white noise due to truncation. Then Gaussian dither was added before quantization in the NCO. After this, the 10 MHz signal was measured again.

\begin{figure}[ht]
	\centering
	\includegraphics[width=8cm]{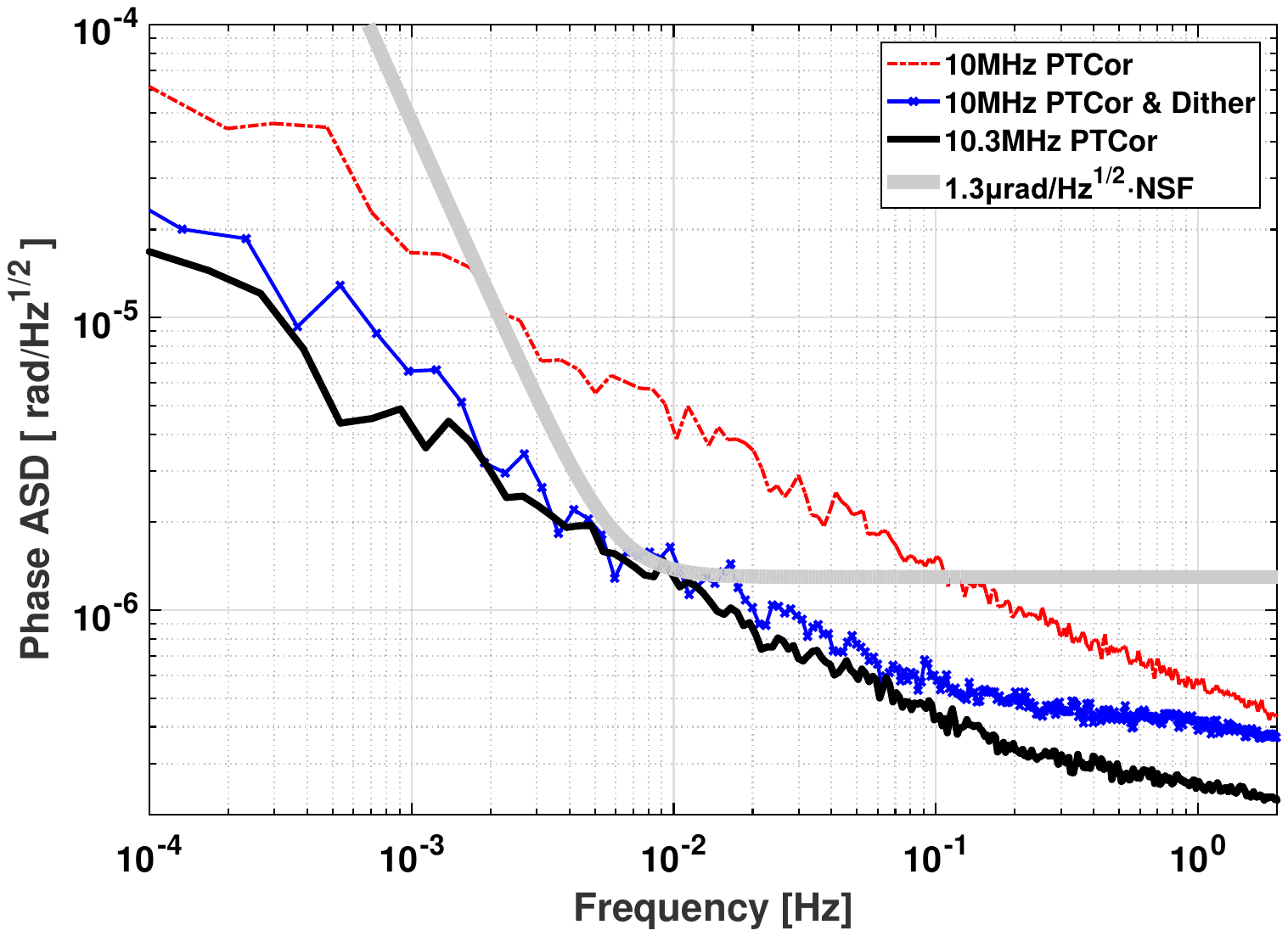}
	\caption{The phase ASD results for signals at different frequencies are presented. The frequencies of the signals are set at 10 MHz and 10.3 MHz, respectively. The dashed red line is influenced by truncation non-white noise, resulting in a high noise floor. Conversely, the dotted blue line is no longer affected by this factor after the introduction of dither, optimizing the noise background to a normal level that meets the project requirements.}
	\label{FIG:10}
\end{figure}

The truncation error exhibits statistical characteristics of non-white noise. Phase truncation results in spurs and aliasing, with remapped spectral lines concentrated around 10 MHz, leading to an increased linewidth of the carrier signal. As shown by the dashed red line in \textcolor{blue}{Fig.}\ref{FIG:10}, the phase noise floor increases as the frequency decreases. The solid gray line represents the phase measurement noise requirement of 1.3 $\rm{\upmu rad/Hz^{1/2}}$ with a noise shape function (NSF) given by $[1+(6~{\rm{mHz}}/f)^4]^{1/2}$. When the signal is at 10.3 MHz, the solid black line indicates that this requirement is met in the frequency band from 0.1 mHz to 1 Hz. The 10 MHz signal is indicated by the dashed red line. From 2 mHz to 0.1 Hz, the noise floor of the phasemeter exceeds the requirement due to the integer multiple.

After adding dither, the 10 MHz signal was measured again. The effect of truncation can be almost mitigated, although additional white noise was introduced above 0.1 Hz as a result of Gaussian dither. At 10 mHz, the phase readout resolution for the 10 MHz signal is 3.9 $\rm{\upmu rad/Hz^{1/2}}$, as indicated by the dashed red line. The phase improved to about 1.3 $\rm{\upmu rad/Hz^{1/2}}$, as indicated by the dotted blue line. Compared to the dashed red line, this optimization of 9.5 dB meets the requirement for phase readout resolution in space gravitational wave detection.

\section{Conclusion}
Driven by FPGA and DSP technologies, the integration and functionality of modern digital phasemeters have significantly improved, endowing them with high-precision measurement and analysis capabilities for dynamic signals. The quantization error, an important noise source in digital circuits, is a deterministic function of the input signal. This paper specifically analyzes the white and non-white noise characteristics associated with the quantization errors of phase truncation in digital phasemeters.

The quantization error can be considered white noise under specific conditions, which power is related to the resolution of quantizer and is uniformly distributed within the Nyquist frequency. However, the non-white noise characteristics of quantization noise occur when the signal frequency and sampling frequency are close to an integer multiple. The spectral lines remapped by aliasing cause the NCO output to deteriorate with an increase in spectral linewidth, resulting in additional low-frequency phase noise. Moreover, the generation of specific artifacts introduces a nonlinear effect on the measured value. To mitigate these effects, Gaussian dither synthesized by a LFSR is added before quantization to smooth the truncation process. Through the analysis, we found that for a 10 MHz signal, the noise floor of the phasemeter exceeds the requirement in the frequency band from 2 mHz to 0.1 Hz due to the integer multiple. After adding dither, the phase noise was optimized by 9.5 dB, meeting the noise requirement of 1.3 $\rm{\upmu rad/Hz^{1/2}} \cdot \rm{NSF}$ from 0.1 mHz to 1 Hz. It is demonstrated that adding dither can effectively suppress the low-frequency phase noise caused by phase truncation.

\bibliographystyle{unsrt}
\bibliography{cas-refs}

\end{document}